\documentclass{svproc}
\usepackage{url}

\usepackage{hyperref}       
\usepackage{url}            
\usepackage{booktabs}       
\usepackage{amsfonts}       

\usepackage{graphicx}
\usepackage{caption}
\usepackage{subcaption}

\usepackage{float}

\usepackage{algorithm}
\usepackage{algpseudocode}
\usepackage[colorinlistoftodos]{todonotes}

\usepackage{amsthm}
\usepackage{amsmath,amssymb,url,longtable,booktabs,multicol,multirow,lineno}

\usepackage[multiple]{footmisc}
\usepackage{tikz}
\usepackage{tikzscale}
\usepackage{xspace}
\usepackage{pgfplots}

\usepackage[multiple]{footmisc}
\usepackage{tikzscale}
\usepackage{xspace}
\theoremstyle{definition}

\begin{document}
\mainmatter              
\title{Locally Weighted Mean Phase Angle (LWMPA) Based Tone Mapping Quality Index (TMQI-3)}
%
%

\author{Inaam Ul Hassan,$^{1\dagger}$ Abdul Haseeb,$^{2\dagger}$ 
Sarwan Ali$^{3\ast}$\\
  {$^{1}$Department of Computer Science, \\ Lahore University of Management Sciences,}{ Lahore, Pakistan}\\
  {$^{2}$Department of Computer Science, CECOS University,}\\
  {Peshawar, Pakistan}\\
  {$^{3}$Department of Computer Science, Georgia State University,}\\
  {Atlanta, GA 30303, USA}\\
  {$^\dagger$ Joint first authors}\\
  {$^\ast$Corresponding author}\\
  {email: 16030050@lums.edu.pk; ahaseebkhan123@gmail.com;}\\
  {sali85@student.gsu.edu}\\
  \institute{}
}

\maketitle              

\begin{abstract}

High Dynamic Range (HDR) images are the ones that contain a greater range of luminosity as compared to the standard images. HDR images have a higher detail and clarity of structure, objects and colour, which the standard images lack. HDR images are useful in capturing scenes that pose high brightness, darker areas, and shadows, etc. An HDR image comprises of multiple narrow-range-exposure images combined into one high-quality image. 
As these HDR images cannot be displayed on standard display devices, the real challenge comes while converting these HDR images to Low dynamic range (LDR) images. The conversion of HDR image to LDR image is performed using Tone-mapped operators (TMOs). This conversion results in the loss of much valuable information in structure, colour, naturalness and exposures. 
The loss of information in the LDR image may not directly be visible to the human eye.
To calculate how good an LDR image is after conversion, various metrics have been proposed previously. 
Some are not noise resilient, some work on separate colour channels (Red, Green, and Blue one by one) and some lack capacity to identify the structure. To deal with this problem, we propose a metric in this paper called the Tone Mapping Quality Index (TMQI-3), which evaluate the quality of the LDR image based on its objective score. TMQI-3 is noise resilient, takes account of structure and naturalness, and works on all three colour channels combined into one luminosity component. This eliminates the need to use multiple metrics at the same time. We compute results for several HDR and LDR images from the literature and show that our quality index metric performs better than the baseline models.  

\end{abstract}
\section{Introduction}
The pictures taken from a camera is a combination of two components, the luminance component and the chrominance component. The visibility of the chrominance component is highly dependant on the intensity of white light presented by the luminance component. \cite{colorLuminance_parraga_1998,gijsenij2011computational,oliva2000diagnostic}. The normal picture consists of shadows and brighter portions. Sometimes due to brighter parts, the details for shadow portions becomes very low, and in worst cases, the objects in that portion of the image become near to invisible. This is termed a Low Dynamic Range (LDR) image~\cite{marnerides2018expandnet}. One way of balancing the brightness and shadow areas, such that both are visible clearly and no detail is missed, is called High Dynamic Range (HDR) Imaging~\cite{Color_Fairchild_2007}. 

According to \cite{lezoray2016high}, when we capture the high contrast image from a camera, either dark images are saturated, or the bright ones are saturated in the output image. This occurs because the sensors in our camera are minimal to capture the luminosity due to shadows or over-brightness. Our conventional cameras cannot provide a proper image, the same as a human eye. The idea of HDR photograph is presented to us. Traditionally for creating an HDR photograph, various LDR images are combined. But the prime issue is that most of the devices are unable or incapable to show HDR images. Therefore, it is important to convert the HDR image to the LDR image without any loss of information. 

The process of creating HDR images is done by capturing multiple shots (3 or more) of various exposures and combining them into one better image. High-dynamic-range imaging (HDRI) \cite{reinhard2010high,nayar2000high} is a technique in which images are processed or captured with a higher amount (or level) of luminosity. This higher dynamic range of luminosity cannot be achieved through standard digital imaging. To capture a HDR image, several narrow range exposed images are taken and combined into one. A limited exposure range mainly results in the loss of important information such as highlights and shadows.
The method of capturing HDR images is quite common and mostly the same in all techniques. Still, the significant issue here is the representation of those High contrast and High dynamic images onto our typical day to day devices which have minimal display powers and are capable enough to display the LDR images only.
Another crucial part in HDR image rendering and displaying is \textit{Tone Mapping}~\cite{mantiuk2008display,Tone_Salih_2012}. 

Tone mapping corresponds to rendering an HDR image to a standard monitor or printing device. This process is carried out by Tone mapping operators (TMO). The need to rendering HDR to LDR is crucial because an HDR image cannot be directly displayed on a standard display or printing device due to its high contrast and colour ratios. 
There are different TMO proposed previously for the conversion of HDR to LDR \cite{ka2016local,ma2015high,ma2014high,Tone_Salih_2012}.
A natural question that arises is regarding the quality of images resulting from applying TMOs on them. Two types of evaluation can be applied to the images, namely subjective evaluation and objective evaluation. Subjective evaluation is done using human eyes, while objective evaluation is done by analyzing the actual structure of the images.


HDR contains a full noticeable scope of luminance and shading of the images~\cite{luminance_Kevin_2010}. However, while converting images from HDR to LDR (using TMOs), there is a possibility of missing essential image structures in the resultant LDR image~\cite{yeganeh2012objective}. It is challenging for human eyes to catch these structural problems in the LDR image (called subjective evaluation). This problem creates room for an objective image quality measure of the tone mapped images so that the overall quality of an image can be measured in detail.

Tone mapping quality index (TMQI) is used to objectively measure the quality of the images (called objective evaluation) converted from HDR to LDR by the TMOs. Previously, different objective quality measure are proposed such as TMQI-1 \cite{yeganeh2012objective}, TMQI-2 \cite{ma2015high}, and Feature Similarity Index for Tone-Mapped Images (FSITM)~\cite{nafchi2014fsitm}. Although FSITM shows better results as compared to TMQI-1 and TMQI-2, the major problem with it is that it works with Red, Blue, and Green channels separately, rather than combined. We propose a new approach, called TMQI-3, that overcomes the limitations of FSITM. Our contributions in this paper are the following:
\begin{enumerate}
    \item We study different tone mapping operator and through experiments we identify the strength and weaknesses of each operator.
    \item We propose an algorithm for objective quality assessment of images called tone-mapped image quality index (TMQI-3).
    \item Our proposed quality index algorithm shows good correlations between a subjective ranking score of the images and an objective ranking score computed using TMQI-3.
\end{enumerate}

\section{Literature Review}
Objective assessment is a common approach to analyse the output of algorithms in many domains such as graph analytics~\cite{ali2021predicting,ahmad2020combinatorial}, protein sequence study~\cite{ali2021spike2vec,ali2021effective,ali2021k,ali2021simpler}, smart grid~\cite{ali2019short,ali2019short2,ali2020fair}, information processing~\cite{ali2021cache}, network security~\cite{ali2020detecting}, and pattern recognition~\cite{ullah2020effect}.
In the images quality assessment domain, in spite of operating on the original HDR pixel value, almost every tone mapping algorithm perform the task on the logarithm of luminance values of HDR pixel.
Olivier Lezoray in \cite{lezoray2016high} proposes to use manifold based ordering in which pixel value is more non-linear rather than a log-luminance curve. The proposed method learns the manifold of HDR image pixels to build a new HDR image representation. That image representation is in the form of an index image associated with an ordering of the HDR pixels’ vectors. It was done in the following manner:

\begin{enumerate}
\item  Ordering of HDR vectors
\item According to the order of first step The Novice depiction of HDR images are performed

\end{enumerate}

Authors in \cite{mantiuk2008modeling} propose a generic tone mapping algorithm that can be used in the black-box analysis of existing TMO, backward compatibility of HDR image compression, and the synthesis of new algorithms that are a combination of existing operators. Their paper says that one model does the estimation of TMOs comprise of tone curve accompanied by spatial modulation function. Moreover, nearly the same image processing technique is used by approximately all TMOs but the selection of parameters varies.

  

Eilertsen et. al. in \cite{eilertsen2017comparative}, authors have done a broad comparison between different TMOs and point out the drawbacks in video TMOs.  
Their prime focus is on the descriptive analysis of the new changes and evolution in tone mapping pipelines. They also devised a new and generic tone mapping algorithm that can best suit the needs of future HDR videos

Kede et. al. In \cite{ma2015high} the authors introduce TMOs that intend to pack high dynamic range (HDR) pictures to Low Dynamic Range (LDR) to envision HDR photos on standard presentations.They have proposed a significantly unique way to deal with outline TMO. Rather than utilizing any predefined efficient computational structure for tone mapping, they specifically explore all pictures in the space of all photographs, hunting down the image that advances an enhanced TMQI. Specifically, the first enhance the two building hinders in TMQI—primary devotion and genuine expectation parts—prompting a TMQI-2 metric. They then propose an iterative calculation that, on the other hand, enhances the auxiliary loyalty and measurable expectation of the following picture. Numerical and subjective tests show that the proposed analysis reliably delivers better quality tone-mapped images, notwithstanding when the most focused TMOs make the underlying pictures of the emphasis. Then, these outcomes likewise approve the predominance of TMQI-2 over TMQI-1.

In \cite{nafchi2014fsitm} the authors suggested a feature similarity index for tone-mapped images (FSITM) system that works on the local phase information of images.  For assessing the tone mapping operator (TMO), the suggested index compares the calculated associated tone-mapped image by utilising the TMO method's output with the locally weighted mean phase angle map of an original high dynamic range (HDR). For experiments, they have taken two sets of images after assessing the results. It shows that the FSITM system outperforms the other tone mapped quality index (TMQI) algorithms. Furthermore, they combine the proposed system FSITM with TMQI, and show better results as compared to typical TMQI's.

Authors in \cite{ka2016local} Introduce a new hybrid method that has been introduced by combining two-hybrid tone mapping operators (local and global operators). Several images are amalgamating into a single HDR which results in enhanced HDR image.  An enhancement map is constructed either with the threshold value or the luminance value of the pixel. Using the enhanced map, the original luminance map is separated from the base layer and detail layer by running bilateral filtering (noise-reducing filter for images). The detail layer is used to enhance the result of global tone mapping. The performance of hybrid tone mapping is then compared to individual local and international operators, and the results show that the hybrid operator gives better performance. 

\section{Proposed Approach}
This section proposes our algorithm, TMQI-3, for objectively evaluating LDR images (performance of TMO's).
In the literature, there are two types of popular models for measuring the quality of images.
\begin{enumerate}
    \item Peak signal-to-noise ratio
    \item Structural Similarity Index Metric (SSIM)
\end{enumerate}
However, the above two quality measures assume that the reference and compared images have the same dynamic range. Since that assumption is not valid in the LDR images, we cannot directly apply these models in our research.

\begin{definition}
TMQI-1 assesses the quality of individual LDR images based on combining an SSIM-motivated structural fidelity measure with the statistical naturalness. The expression for TMQI-1 is given in Equation~\eqref{eq_tmqi_first}.
\end{definition}

\begin{equation}\label{eq_tmqi_first}
    \text{TMQI-1} (x,y) = a[S(x,y)a + (1-a) N(y)]B
\end{equation}
$S$ represents structural fidelity, $N$ represents statistical naturalness, $x$ represents HDR image, and $y$ represents the LDR image. The parameters $a$ and $B$ determine the sensitivity. The range of the parameter $a$ is the following:
\begin{equation}
    0 \leq a \leq 1
\end{equation}
\begin{remark}
Note that the parameters structural fidelity and statistical naturalness are upper bounded by $1$. Therefore, TMQI-1 is also upper bounded by $1$.
\end{remark}

TMQI-1 was the first approach for measuring the LDR image's quality across the dynamic range to the best of our knowledge. 
It provides a better assessment for the LDR images (compared to the traditional methods discussed above) as a result of applying TMO on the HDR. 
However, TMQI-1 have some limitations such as
\begin{enumerate}
    \item It can only be applied to greyscale images. However, most of the HDR images nowadays are in colour.
    \item The statistical naturalness measure used in the TMQI-1 is based on the intensity statistics only. However, many sophisticated statistical models in the literature can also capture other properties of the image, such as structural regularities in space, scale, and orientation.
\end{enumerate}

\begin{remark}
The problem of TMQI-1 only working with the greyscale images can be solved by applying the TMQI-1 on each colour channel separately. Although this may allow TMQI-1 to work on colour images, its performance will not be outstanding~\cite{yeganeh2012objective}.
\end{remark}

We have proposed a new approach in which we have combined $3$ existing techniques differently to achieve better results in terms of subjective evaluation of both TMQI-2 \cite{ma2015high}, and FSITM \cite{nafchi2014fsitm}. The approaches which we have used as a reference for our research are the following.

Our proposed method (TMQI-3) take into account the following properties of the images.
\begin{enumerate}
    \item Structural fidelity
    \item Statistical naturalness
    \item It also makes use of the mean phase angle of local weights
\end{enumerate}
By combining the properties mentioned above, we could get a better objective score for the input images.

We will now discuss all the above properties one by one.

\subsection{Improved Structural Fidelity} 
The structural fidelity~\cite{ma2015high} of TMQI-1 can be computed using a sliding window across the whole image. This process results in a quality map and hence preserving the local structural detail of the image.
Local structural fidelity measure given by TMQI-1 is given in Equation. \eqref{eq_s_local}.

\begin{equation}\label{eq_s_local}
    S_{local}(x,y) = \frac{2 \hat{\sigma}_x \hat{\sigma}_y + C_1}{\hat{\sigma}_x^2 \hat{\sigma}_y^2 + C1} .  \frac{\sigma_{xy} + C_2}{\sigma_x \sigma_y + C2}
\end{equation}


where $\sigma_x$ and $\sigma_y$ denote the local standard deviations (std), respectively. The $\sigma_{xy}$  denotes the covariance between two corresponding patches. The two (positive) constant terms C1 and C2 are used to avoid any possible instability. 
Overall structural fidelity is calculated using Equation. Equation~\eqref{eq_s_x_y}.
\begin{equation}\label{eq_s_x_y}
    S(X,Y) = \frac{1}{M} \sum_{i = 1}^M S_{local}(x_i,y_i)
\end{equation}


The updates in structural fidelity are done through the gradient descent method stated in Equation. \eqref{eq_y_hat}.
\begin{equation}\label{eq_y_hat}
    \hat{Y}_k = Y_k + \lambda \nabla_Y S(X,Y)\vert_{Y = Y_k}
\end{equation}


Where $Y_k$ is the image resulting from k iteration and $\lambda$ is the step size. This works as the contrast visibility model for the local luminance model. 

\subsection{Improved Statistical Naturalness}
Model for statistical naturalness \cite{ma2015high} proposed in TMQI-1 is given in Equation \eqref{eq_n_y}.
\begin{equation}\label{eq_n_y}
    N(Y) = \frac{1}{K} P_m P_d
\end{equation}
where $P_m$ represents gaussian density function while $P_d$ represents beta density function.	
However, Equations~\eqref{eq_n_y} has the following limitations.
\begin{enumerate}
    \item Gaussian density function and beta density function are considered independent of image content, which may not entirely be true. 
    \item Model for statistical naturalness is derived from high-quality images while having no information regarding how an unnatural image may look.
\end{enumerate}

The updates above can be abstractly defined by the equations below:

\begin{equation}
    Pm = \frac{1}{\sqrt{2 \pi \theta_1}} \int_{-\infty}^{\mu} exp \Big[ - \frac{(t-T_1)^2}{2 \theta^2_1} \Big]   dt \mu \leq \mu_e
\end{equation}

\begin{equation}
    Pm = \frac{1}{\sqrt{2 \pi \theta_2}} \int_{-\infty}^{2\mu_r - \mu} exp \Big[ - \frac{(t-T_2)^2}{2 \theta^2_2} \Big]   dt \mu > \mu_e
\end{equation}

\begin{equation}
    Pd = \frac{1}{\sqrt{2 \pi \theta_3}} \int_{-\infty}^{\sigma} exp \Big[ - \frac{(t-T_3)^2}{2 \theta^2_3} \Big]   dt \sigma \leq \sigma_e
\end{equation}

\begin{equation}
    Pd = \frac{1}{\sqrt{2 \pi \theta_4}} \int_{-\infty}^{2\sigma_r - \sigma} exp \Big[ - \frac{(t-T_4)^2}{2 \theta^2_4} \Big]   dt \sigma > \sigma_e
\end{equation}

\begin{remark}
Note that the acceptable luminance changes saturate at both small and large luminance levels without significantly tampering with the image visual naturalness \cite{ma2015high}.
\end{remark}

\subsection{Use of Mean Phase Angle of Local Weights}
An image quality measure based on the local phase information of an image is proposed in~\cite{nafchi2014fsitm}. Their model is noise independent; hence no parameter is required for noise estimation.
\begin{remark}
Note that multiple methods have already been proposed in the literature related to the quality assessment of images using the phase. information~\cite{nafchi2014fsitm,zhang2011fsim,oppenheim1981importance,concetta1988feature,hassen2013image,saha2013perceptual}.
\end{remark}
One drawback of the methods proposed in the literature that consider phase information is that their results for evaluating tone-mapped images are not reliable compared to other famous quality assessment metrics like SSIM. 
The Locally Weighted Mean Phase Angle (LWMPA) is robust to noise and is computed using the following expression according to~\cite{nafchi2014fsitm}.

\begin{equation}\label{eq_ph_val}
    ph(x) = \text{arctan} \ 2 [\sum_{\rho,r}^{} e_{\rho r} (x), \sum_{\rho,r}^{} o_{\rho r} (x)]
\end{equation}
where $\rho$ represents scale and $r$ represents orientation of the image~\cite{nafchi2014fsitm}.
The values for the pixels ($v_{ph}$) of $ph(x)$ are following
\begin{equation}\label{eq_pixel_val}
    \frac{- \pi}{2} \leq v_{ph} \leq \frac{+ \pi}{2}
\end{equation}
The term $\frac{- \pi}{2}$ in Equation~\eqref{eq_pixel_val} represents a dark line while $\frac{+ \pi}{2}$ represents a bright line. The pixels of $ph(x)$ takes value $0$ for the steps. For further detail, readers are referred to \cite{kovesi2002edges,nafchi2014fsitm}.


The ph(x) in Equation~\eqref{eq_ph_val} provides a good representation of image features. This representation includes the edges of the objects within the image and the shapes of those objects. From Equation~\eqref{eq_pixel_val}, we know that ph(x) represents both dark and bright lines. Therefore, it can be used to identify colours within the image. This colour detection is a useful property for a TMO. 
\begin{remark}
Note that the LWMPA has the property to ignore the noise from the image, which is not the case with the TMQI-1 that uses phase derived features.
\end{remark}


FSITM works on Locally Weighted Mean Phase Angle (LWMPA). It works on separate channels of Red, Blue, and Green. Not all of them combined. This is a significant issue that we noticed in FSITM. The image should be judged as a whole, not on separate channels only and according to the human eye's sensitivity. Combining the locally weighted mean phase angle with structural fidelity and statistical naturalness requires it to be represented in luminance. The reason is that the structural fidelity and statistical naturalness work upon the luminance of the image. So to combine the LWMPA with others, its quality index score should be mapped for luminance components, too, since the human eye is much more responsive to the luminance rather than RGB separately. To solve the problem that we identified in FSITM, we used the YUV model. The equation of ‘Y’ is used to connect Red, Green and Blue to luminance. We used this equation for the following reasons. 
\begin{enumerate}
    \item The conversion of RGB to Y requires just a linear transform, which is very easy to do and cheap to compute numerically
    \item ‘Y’ is perceived as brightness, which is more sensitive to the human eye. Luminance gives the measure of the amount of energy an observer perceives from the light source. That is why separate RGB sensitivity values are selected in Y. 
\end{enumerate}

Its equation for Y is the following:

\begin{equation}
    Y = 0.299*R + 0.587*G + 0.114*B
\end{equation}

TMQI-1, TMQI-2, and FSITM work on different methods to evaluate the image. Structural fidelity and statistical naturalness are considered in TMQI-2, while LWMPA is considered in FSITM. Separately, these methods are suitable to access some types of image features (but not an image as a whole). So they should be combined into one quality assessment model. For this purpose, we combine all of these to design our TMQI-3 and give equal weight to each component, as given below:

\begin{equation} \label{weightsEqs}
    Q = (1/3)* N + (1/3)* F + (1/3)* L
\end{equation}
Where N, F, L denote statistical naturalness, structural fidelity, and LWMPA, respectively.

The improvements in structural fidelity and statistical naturalness are somehow provided in TMQI-2, but it lacks noise resilience like in SSIM and other matrices. However, they give better results due to gradient ascent method infidelity and solving parameter optimization problems for point-wise intensity transformation in statistical naturalness. Further noise reduction can be made using the phase angle map of locally weighted means because it is robust and has error resilience. By combining improvements in structural fidelity and statistical naturalness with phase angle map, we would be improving the quality and hence enhance our quality index.

Our metric takes following $3$ inputs. 
\begin{enumerate}
    \item An HDR image is used as a reference
    \item The LDR image being compared (either colour or greyscale image with its dynamic range equal to 255)
    \item A local window for statistics
\end{enumerate}
The default window for our approach is gaussian. The quality metric first computes the structural fidelity and statistical naturalness of the image using the method proposed by \cite{ma2015high}. 
Then it computes the quality index for RGB separately by the method provided in \cite{nafchi2014fsitm}. The main issue comes at the point of combining these three. For this purpose, RGB quality index values are mapped to luminance quality index values, as stated in the previous paragraphs. They are combined according to the sensitivity of RGB values to human eyes. Then after that, these three are combined with equal weights given to each.

ombining these three different approaches comes from the fact that each method checks the image from a different perspective. The structural fidelity focuses upon the visibility of image details which further depends upon the sampling density of the image, the distance between the image and the observer, the resolution of the display, and the perceptual capability of the observer's visual system. For the statistical naturalness, the studies in \cite{yeganeh2012objective} show that among all attributes such as brightness, contrast, colour reproduction, visibility, and reproduction of details, brightness and contrast have more correlation with perceived naturalness. The real challenge of combining these three together was due to the LWMPA from FSITM. The image is not observed based on chrominance, luminance (brightness) or contrast, but based on the three primary colours only, that is, Red, Green and Blue, and the quantities of these three in an image. LWMPA is not susceptible to noise or error, which makes it more robust in terms of quality. Combining these three results in a Quality matrix with the capability of luminance, contrast, Structure, Naturalness, and the phases of the image. 

We would effectively distinguish the visible and invisible local contrast in images and provide a good representation of image features indicating the changes in colours and dark or bright lines.

\section{Results and Discussion}
In this section, we first give information about the dataset and other implementation detail. Then we show our results and discuss the behaviour of our proposed algorithm is compared to the baselines.

\subsection{Experimental Setup}
To evaluate our metric, we used the dataset proposed in \cite{yeganeh2012objective}. It contains 15 HDR images, with each image having its associated 8 LDR images as well. The LDR images are subjectively scored from the range of 1 to 8. A subjective score of 1 means the image is best converted from HDR to LDR, and a score of 8 means the worst conversion. 
The subjective scores were obtained based on an assessment of $20$ individuals.
Implementation of our algorithm is done in Matlab, and experiments are performed on Core I$3$ $2.4$ GHz with $8$ GB of RAM.


The evaluation metric that we are using is Kendall’s rank-order correlation coefficient (KRCC) \cite{yeganeh2012objective}. KRCC is a non-parametric rank correlation metric whose formula is the following:
\begin{equation}
    KRCC = \frac{N_c - N_d}{\frac{1}{2} N(N-1)}
\end{equation}
Where $N_c$ is the number of consistent rank order (concordant), and $N_d$ is the number of inconsistent rank order (discordant) pairs in the data set.

\subsection{Kendall Correlation Based Results}
Kendall's correlation coefficient was run on the quality indices generated for each method (TMQI-3, TMQI-2, TMQI-1, FSITM), and the results are shown in Table \ref{tbl_main_results}. We can observe that the TMQI-3 process is comparable (also better for some images) to the TMQI-2 approach. 
In reference to TMQI-1, the TMQI-3 performs better in many cases. In FSITM, TMQI-3 is better for image set 7 and comparable for image set 1, 6, 11, and 15.

\begin{table}[h!]
    \centering
    \begin{tabular}{ccccc}
    \hline
       Image Set & TMQI-3 & TMQI-2 & TMQI-1 & FSITM \\
       \hline \hline
       1 & 0.7857 & 0.7857 & 0.3571 & 0.7857 \\
       2 & 0.3571 & 0.2857 & 0.6429 & 0.7143 \\
       3 & 0.5714 & 0.5714 & 0.6429 & 0.7857 \\
       4 & 0.5000 & 0.5000 & 0.7143 & 0.7857 \\
       5 & 0.5000 & 0.5000 & 0.6429 & 0.6429 \\
       6 & 0.7857 & 0.7143 & 0.7143 & 0.7857 \\
       7 & 0.7857 & 0.7143 & 0.5714 & 0.7143 \\
       8 & 0.5714 & 0.5000 & 0.5714 & 0.6429 \\
       9 & 0.7143 & 0.7143 & 0.5714 & 0.8571 \\
       10 & 0.7857 & 0.7857 & 0.8571 & 0.8571 \\
       11 & 0.7143 & 0.7143 & 0.7143 & 0.7143 \\
       12 & 0.4286 & 0.4286 & 0.5714 & 0.5714 \\
       13 & 0.6071 & 0.6071 & 0.5357 & 0.6786 \\
       14 & 0.5714 & 0.5714 & 0.6429 & 0.6429 \\
       15 & 0.7857 & 0.7857 & 0.7857 & 0.7857 \\
       \hline \hline 
       Average & 0.6309 & 0.6119 & 0.6357 & 0.7309 \\
       Min & 0.3571 & 0.2857 & 0.3571 & 0.5714 \\
       Max & 0.7857 & 0.7857 & 0.8571 & 0.8571 \\
       Std & 0.1402 & 0.1445 & 0.1138 & 0.0812 \\
       \hline
    \end{tabular}
    \caption{
KRCC values between subjective score and different TMQI's score (higher score is better). For each of the 15 HDR images, we have 8 LDR images. We computed each LDR image's objective score separately and then reported their average in this table (for all techniques). The Average, Minimum, Maximum, and Standard Deviation values are also reported.}
    \label{tbl_main_results}
\end{table}



We can argue that although there is not any clear winner from Table~\ref{tbl_main_results}. FSITM looks to be better than the other methods. However, authors of the FSITM in~\cite{nafchi2014fsitm} argues that different TMO algorithms perform differently on additional HDR images. Their behaviour depends on the (type of) HDR image to be converted. From this uncertainty of the TMO's, we can conclude that the best.
TMO approach must be found for each case (no TMO can be generalized on all HDR images).


\subsection{Visual Results on LDR Images}
In this section, we explore the visual effect of different LDR images generated  from different TMO's \cite{nafchi2014fsitm}.

\subsubsection{Indoor House Images:}
\begin{figure}[h!]
\minipage{0.33\textwidth}
  \includegraphics[width=\linewidth]{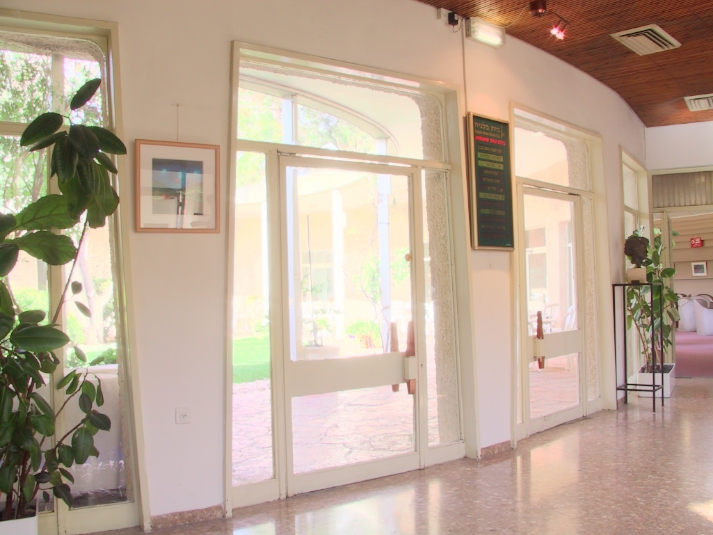}
  \caption*{(a)}
\endminipage\hfill
\minipage{0.33\textwidth}
  \includegraphics[width=\linewidth]{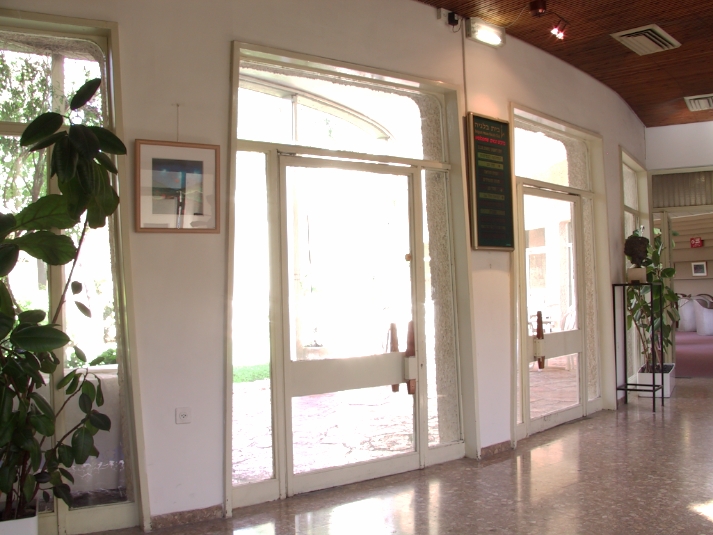}
  \caption*{(b)}
\endminipage\hfill
\minipage{0.33\textwidth}%
  \includegraphics[width=\linewidth]{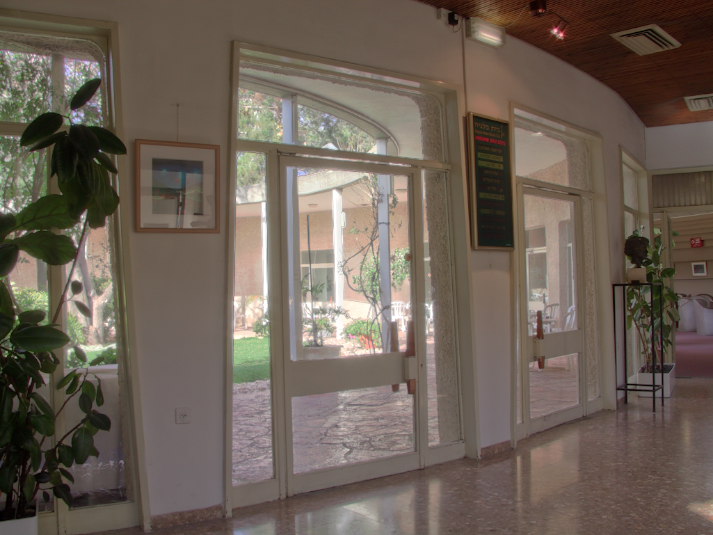}
  \caption*{(c)}
\endminipage
\caption{Indoor house LDR images created using different TMOs~\cite{yeganeh2012objective}. The subjective and objective scores for each image are given in Table~\ref{tbl_indoor_house}.}
  \label{fig_result_1}
\end{figure}

These first set of LDR images in Figure \ref{fig_result_1} are of indoor houses generated from their corresponding HDR images. Referring to the Table~\ref{tbl_indoor_house} we can identify that the worst subjective score is given to the Figure \ref{fig_result_1}(b) (which is 5.95). 
\begin{remark}
A higher subjective score (max 8) refers to a bad conversion from HDR to LDR, while a lower subjective score (minimum 1) refers to better conversion.
\end{remark}
We can observe that the best subjective score is of the Figure \ref{fig_result_1}(c). For each resultant Figure TMQI-3, TMQI-2, TMQI-1, and FSITM are run to produce objective scores. The results can be seen in Table~\ref{tbl_indoor_house}. We can observe the correspondence between the objective and subjective scores in the associated table. 
\begin{remark}
The objective score of a metric should be low if the subjective score is high and high if the subjective score is low. 
\end{remark}
The aim of FSITM scores for Figure~\ref{fig_result_1} (b) should be less than that of Figure~\ref{fig_result_1} (a) because Figure~\ref{fig_result_1} (b) have a higher subjective score. However, in reality, the FSITM is giving the opposite of it. The same behavior is observed with TMQI-1 regarding Figure~\ref{fig_result_1} (b) and Figure~\ref{fig_result_1} (a). The TMQI-2 and TMQI-3 perform as expected, giving a lower objective score to the Figure with a higher subjective score and vice versa. 

\begin{table}[h!]
    \centering
    \begin{tabular}{ccccccc}
    \hline
    & & \multicolumn{4}{c}{Objective scores} \\
    \cline{3-6}
    Figure & Subjective Score & TMQI-3 & TMQI-2 & TMQI-1 & FSITM \\
    \hline \hline
    Fig. \ref{fig_result_1} (a) & 4.4 & 0.5484 & 0.4106 & 0.8016 & 0.8077  \\
    Fig. \ref{fig_result_1} (b) & 5.95 & 0.5194 & 0.3814 & 0.8799 & 0.8311\\
    Fig. \ref{fig_result_1} (c) & 2 & 0.6645 & 0.5770 & 0.9191 & 0.8774\\
    
    \hline
    \end{tabular}
    \caption{Subjective and Objective score for different methods on indoor house LDR image.}
    \label{tbl_indoor_house}
\end{table}

\subsubsection{Open Area Wide Shots Images:}

\begin{figure}[h!]
\minipage{0.33\textwidth}
  \includegraphics[width=\linewidth]{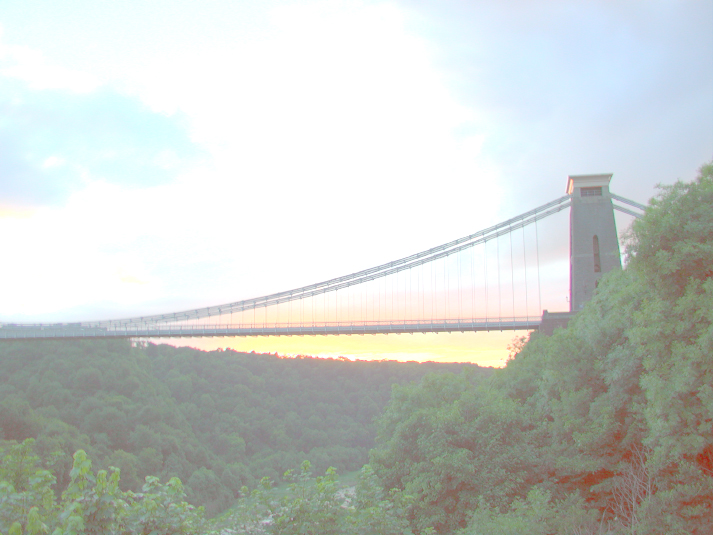}
  \caption*{(a)}
\endminipage\hfill
\minipage{0.33\textwidth}
  \includegraphics[width=\linewidth]{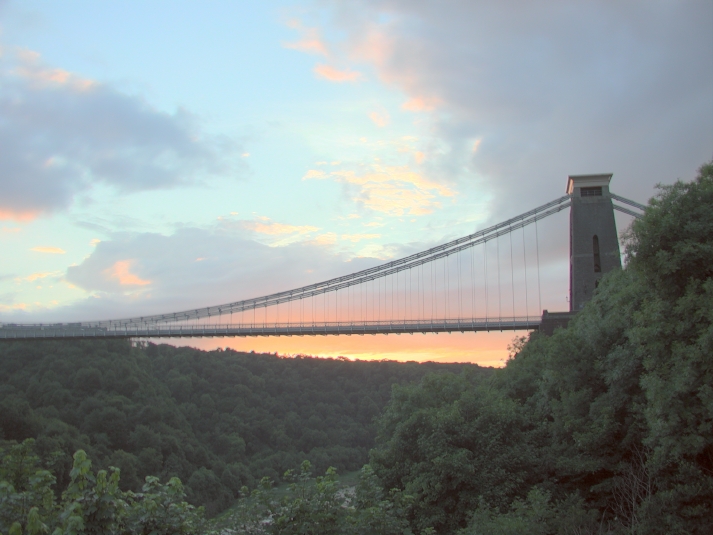}
  \caption*{(b)}
\endminipage\hfill
\minipage{0.33\textwidth}%
  \includegraphics[width=\linewidth]{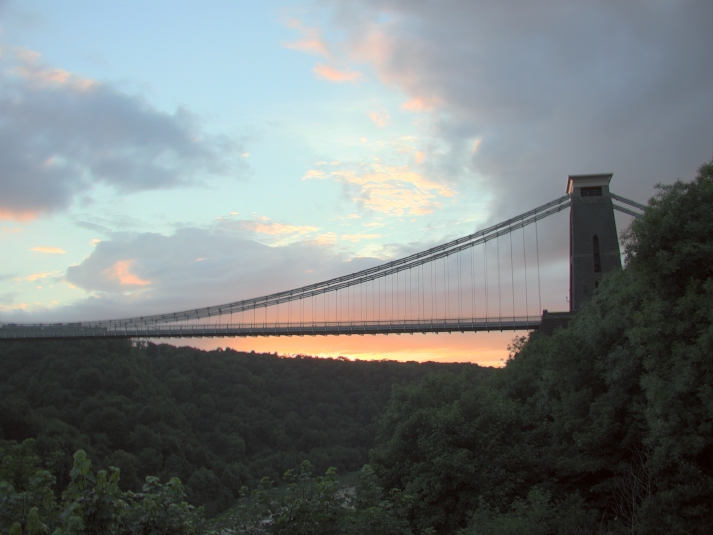}
  \caption*{(c)}
\endminipage
\caption{Open area wide shots LDR images created using different TMOs~\cite{yeganeh2012objective}. The subjective and objective scores for each image are given in Table~\ref{tbl_open_area_wide_shots}.}
  \label{fig_result_2}
\end{figure}

Figure \ref{fig_result_2} shows the LDR images of open area-wide computed by different TMOs from the corresponding HDR image. The Figure \ref{fig_result_2} (a) has the highest subjective score, Figure \ref{fig_result_2} (b) has the lowest and Figure \ref{fig_result_2} (c) is in the middle. The Figure \ref{fig_result_2} (a) has the highest score because of its over brightness and unclear image quality, and Figure \ref{fig_result_2} (b) has the lowest score because of its clear foreground and background. (As stated earlier, a lower subjective score means a better image, and a higher subjective score means the worst image.) The corresponding objective scores of the metrics should be inversely related to the subjective scores of each image, respectively. In this case, all the metrics provide the right objective scores. The objective score and the subjective scores for the Figures are shown in Table~\ref{tbl_open_area_wide_shots}


\begin{table}[h!]
    \centering
    \begin{tabular}{ccccccc}
    \hline
    & & \multicolumn{4}{c}{Objective scores} \\
    \cline{3-6}
    Figure & Subjective Score & TMQI-3 & TMQI-2 & TMQI-1 & FSITM \\
    \hline \hline
    Fig. \ref{fig_result_2} (a) & 7.1 & 0.4547 & 0.3207 & 0.7004 & 0.6987 \\
    Fig. \ref{fig_result_2} (b) & 3.65 & 0.5774 & 0.4732 & 0.8328 & 0.8155 \\
    Fig. \ref{fig_result_2} (c) & 4.75 & 0.8637 & 0.9008 & 0.8589 & 0.8272  \\
    \hline
    \end{tabular}
    \caption{Subjective and Objective score for different methods on open area wide shots LDR image.}
    \label{tbl_open_area_wide_shots}
\end{table}

\subsubsection{Main Object Upfront Images:}

\begin{figure}[h!]
\minipage{0.20\textwidth}
  \includegraphics[width=\linewidth]{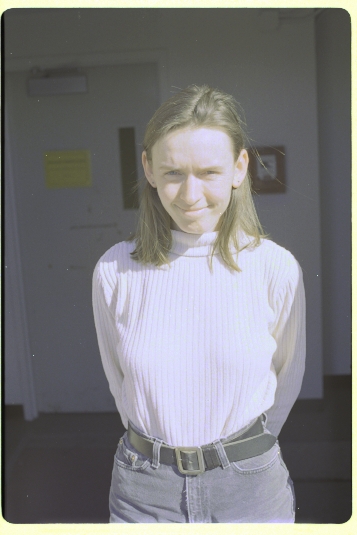}
  \caption*{(a)}
\endminipage\hfill
\minipage{0.20\textwidth}
  \includegraphics[width=\linewidth]{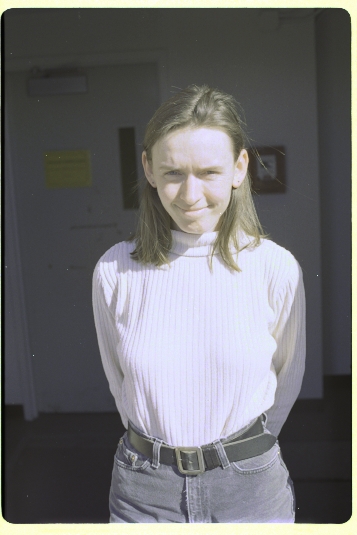}
  \caption*{(b)}
\endminipage\hfill
\minipage{0.20\textwidth}%
  \includegraphics[width=\linewidth]{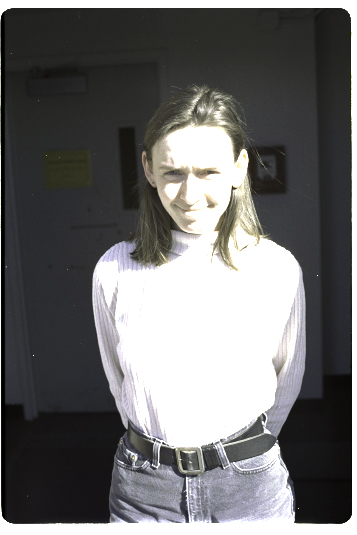}
  \caption*{(c)}
\endminipage
\caption{Main object upfront LDR images created using different TMOs~\cite{yeganeh2012objective}. The subjective and objective scores for each image are given in Table~\ref{tbl_Main_object_upfront}.}
  \label{fig_result_3}
\end{figure}

Figure \ref{fig_result_3} shows the LDR images produced from HDR image from different TMOs. Subjectively the best Figure is \ref{fig_result_3} (b) and the worst Figure is \ref{fig_result_3} (c) (lowest and highest subjective scores respectively). The objective score and the subjective scores for the images are shown in Table~\ref{tbl_Main_object_upfront}. We can notice that the good image has a higher objective score, and the bad image have a lesser objective score in all TMQI-3, TMQI-2, TMQI-1 and FSITM metrics. Thus they all perform well on the images where the main object upfront is obvious than the background. The only difference that can be noticed is that FSITM and TMQI-1 were not able to provide a larger objective-score gap for Figure \ref{fig_result_3} (b) and \ref{fig_result_3} (c). The difference is very minute where it should have been greater as the subjective score of image \ref{fig_result_3} (c) is much greater than that of Figure \ref{fig_result_3} (b). This is not the case with TMQI-2 and TMQI-3. They both provide better differences in objective scores in correspondence to their subjective scores.



\begin{table}[h!]
    \centering
    \begin{tabular}{ccccccc}
    \hline
    & & \multicolumn{4}{c}{Objective scores} \\
    \cline{3-6}
    Figure & Subjective Score & TMQI-3 & TMQI-2 & TMQI-1 & FSITM \\
    \hline \hline
    Fig. \ref{fig_result_3} (a) & 1.65 & 0.7672 & 0.7337 & 0.9363 & 0.8862 \\
    Fig. \ref{fig_result_3} (b) & 1.45 & 0.8801 & 0.9051 & 0.9475 & 0.8926  \\
    Fig. \ref{fig_result_3} (c) & 5.6 & 0.6199 & 0.5544 & 0.9119 & 0.8332 \\
    \hline
    \end{tabular}
    \caption{Subjective and Objective score for different methods on Main object upfront LDR image.}
    \label{tbl_Main_object_upfront}
\end{table}

\subsubsection{Indoor with background scenery:}

\begin{figure}[h!]
\minipage{0.33\textwidth}
  \includegraphics[width=\linewidth]{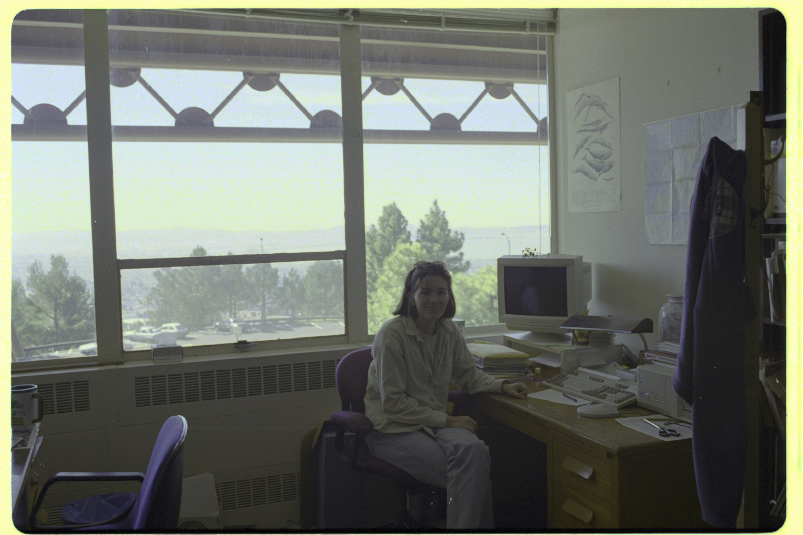}
  \caption*{(a)}
\endminipage\hfill
\minipage{0.33\textwidth}
  \includegraphics[width=\linewidth]{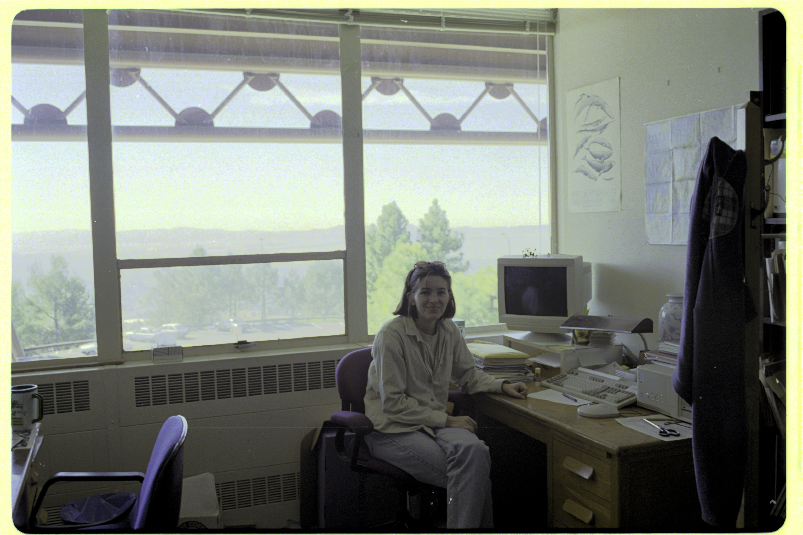}
  \caption*{(b)}
\endminipage\hfill
\minipage{0.33\textwidth}%
  \includegraphics[width=\linewidth]{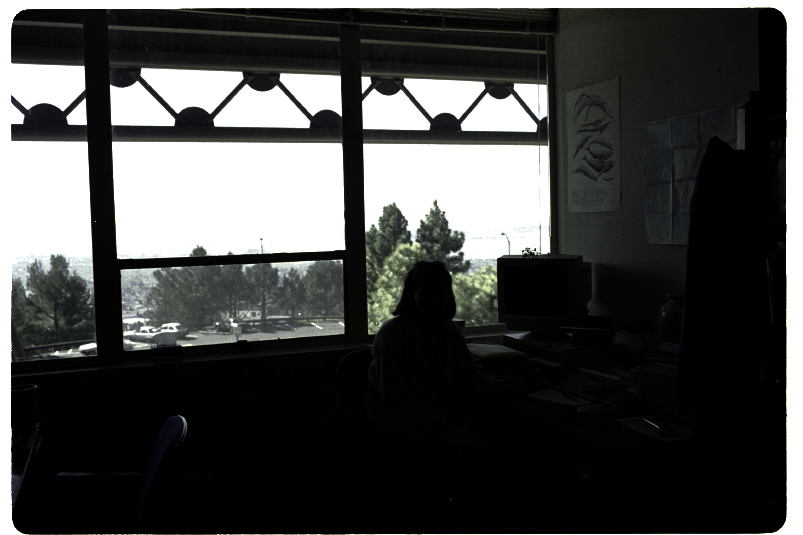}
  \caption*{(c)}
\endminipage
\caption{Indoor with background scenery LDR images created using different TMOs~\cite{yeganeh2012objective}. The subjective and objective scores for each image are given in Table~\ref{tbl_Dull_out_phenomenon}.}
  \label{fig_result_4}
\end{figure}

Figure \ref{fig_result_4} shows the LDR images produced from HDR image by different TMOs. The results are given in Table \ref{tbl_Dull_out_phenomenon}. The best image subjectively so far is image \ref{fig_result_4} (b) and the worst is image \ref{fig_result_4} (c) (scores 1.55 and 5.65 respectively). The objective score from TMQI-3, TMQI-2, TMQI-1, and FSITM are calculated on each image. The objective score of a metric should be low if the subjective score is high and high if the subjective score is low. The FSITM objective scores for image (b) should be less than that of image (a) because the image (b) have a higher subjective score. But in actual FSITM is giving the opposite of it. The same is with TMQI-1 regarding figures (b) and (a) in this case. TMQI-2 and TMQI-3 perform as expected, giving a lower objective score to the image with a higher subjective score and giving a higher objective score to the image with a lower subjective score.


\begin{table}[h!]
    \centering
    \begin{tabular}{ccccccc}
    \hline
    & & \multicolumn{4}{c}{Objective scores} \\
    \cline{3-6}
    Figure & Subjective Score & TMQI-3 & TMQI-2 & TMQI-1 & FSITM-TMQI \\
    \hline \hline
    Fig. \ref{fig_result_4} (a) & 1.55 & 0.8592 & 0.8704 & 0.9476 & 0.8942  \\
    Fig. \ref{fig_result_4} (b) & 3.9 & 0.8316 & 0.8166 & 0.9548 & 0.9050 \\
    Fig. \ref{fig_result_4} (c) & 5.65 & 0.3785 & 0.2142 & 0.8766 & 0.7484 \\
    \hline
    \end{tabular}
    \caption{Subjective and Objective score for different methods on Indoor with background scenery LDR image.}
    \label{tbl_Dull_out_phenomenon}
\end{table}

\subsubsection{Outdoor Natural Scenery Images:}
\begin{figure}[h!]
\minipage{0.33\textwidth}
  \includegraphics[width=\linewidth]{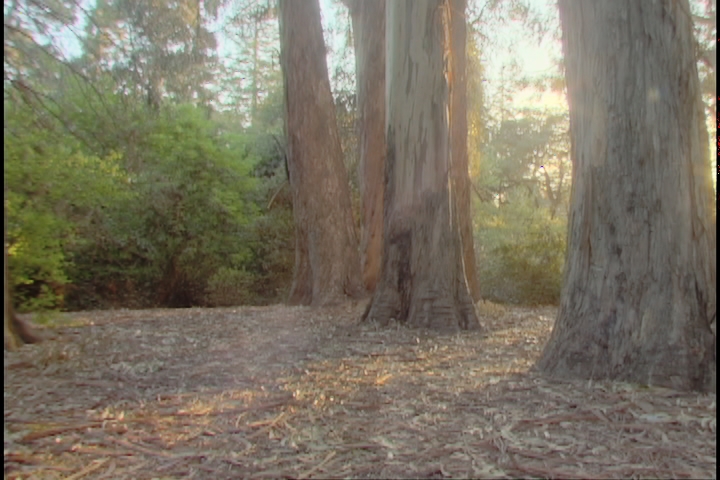}
  \caption*{(a)}
\endminipage\hfill
\minipage{0.33\textwidth}
  \includegraphics[width=\linewidth]{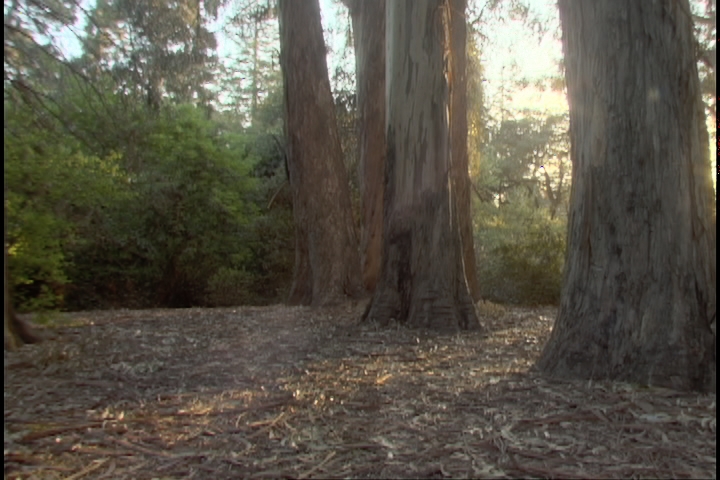}
  \caption*{(b)}
\endminipage\hfill
\minipage{0.33\textwidth}%
  \includegraphics[width=\linewidth]{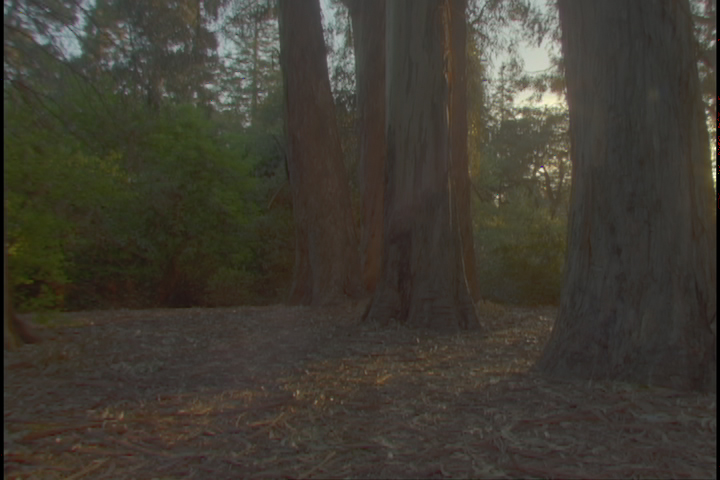}
  \caption*{(c)}
\endminipage
\caption{Outdoor natural scenery created using different TMO's~\cite{yeganeh2012objective}. The subjective and objective scores for each image are given in Table~\ref{tbl_Outdoor_natural_scenary}.}
  \label{fig_result_5}
\end{figure}

This set of LDR images in Figure \ref{fig_result_5} are of outdoor natural scenery generated produced from their respective HDR image through various TMOs. The results are given in Table \ref{tbl_Outdoor_natural_scenary}. It can be observed that the best image (with lowest subjective score) is Figure~\ref{fig_result_5} (a) and worst is Figure~\ref{fig_result_5} (c). Their respective subjective scores are 3.65 and 7.1. It can also be observed that TMQI-3, TMQI-2, TMQI-1, and FSITM all produced the right results. The only difference that can be noticed is that FSITM and TMQI-1 could not provide a larger objective-score gap for each image. Whereas TMQI-2 and TMQI-3 were able to provide better objective score differences with respect to the corresponding subjective scores.


\begin{table}[h!]
    \centering
    \begin{tabular}{ccccccc}
    \hline
    & & \multicolumn{4}{c}{Objective scores} \\
    \cline{3-6}
    Figure & Subjective Score & TMQI-3 & TMQI-2 & TMQI-1 & FSITM \\
    \hline \hline
    Fig. \ref{fig_result_5} (a) & 3.65 & 0.8260 & 0.7768 & 0.9487 & 0.9332 \\
    Fig. \ref{fig_result_5} (b) & 4.75 & 0.7153 & 0.6142 & 0.8952 & 0.9035  \\
    Fig. \ref{fig_result_5} (c) & 7.1 & 0.4975 & 0.3156 & 0.8008 & 0.8167 \\
    \hline
    \end{tabular}
    \caption{Subjective and Objective score for different methods on Outdoor natural scenery LDR image.}
    \label{tbl_Outdoor_natural_scenary}
\end{table}

\section{Conclusion}
We have proposed an objective quality index, called the Locally weighted mean phase angle based tone mapping quality index (TMQI-3), which is based upon the combination of three basic properties namely statistical naturalness, structural fidelity, and locally weighted mean phase angle. These all three properties were used separately in the literature. In this paper, we integrate all of them in one quality index because they can provide better results (when combined) while objectively evaluating an LDR image. The results are seemingly accurate and better than those offered by previous quality matrices.  
Our metric is noise resilient, includes structure and naturalness of image, and and provide better objective scores.
In future, we will focus on developing a better weights assessment strategy for each individual component based on the sensitivity and structure of each image separately. 

\bibliographystyle{plain}
\bibliography{tmqi}

\end{document}